\documentclass{article}
\usepackage{amsmath}
\usepackage{amsthm}
\usepackage{bm}
\usepackage{tikz}
    \usetikzlibrary{calc}
\usepackage{graphicx}
\usepackage[tight,footnotesize]{subfigure}
\usepackage{epstopdf}
\usepackage{booktabs}
\usepackage{threeparttable}
\usepackage{float}
\graphicspath{{./figures/}}
\begin{document}

\title{New method of averaging diffeomorphisms based on Jacobian determinant and curl vector}
\author{Xi Chen, Guojun Liao \\
  \multicolumn{1}{p{.7\textwidth}}{\centering\emph{\scriptsize{Department of Mathematics, \\University of Texas at Arlington, Arlington, Texas 76019, \textsc{USA}\\liao@uta.edu}}}}
%\address{Department of Mathematics, University of Texas at Arlington, Arlington, Texas 76019, \textsc{USA}}
\date{}
\maketitle

\begin{abstract}
  Averaging diffeomorphisms is a challenging problem, and it has great applications in areas like medical image atlases. The simple Euclidean average can neither guarantee the averaged transformation is a diffeomorphism, nor get reasonable result when there is a local rotation. The goal of this paper is to propose a new approach to averaging diffeomorphisms based on the Jacobian determinant and the curl vector of the diffeomorphisms. Instead of averaging the diffeomorphisms directly, we average the Jacobian determinants and the curl vectors, and then construct a diffeomorphism based on the averaged Jacobian determinant and averaged curl vector as the average of diffeomorphisms. Numerical examples with convincible results are presented to demonstrate the method.
\end{abstract}
{\bf Keywords: averaging diffeomorphisms, Jacobian determinant, curl vector, construction of diffeomorphism}
\section{Introduction}
Numerical construction of differentiable and invertible transformations is an interesting and challenging problem. In \cite{Liao2009}, two methods are formulated based on a $div-curl-ode$ system or $div-curl$ system only. The latter is demonstrated in numerical examples to accurately reconstruct 2D and 3D diffeomorphisms using their divergence and curl vector.
%The current paper deals with a much more challenging problem: Numerically construct a %diffeomorphism with prescribed Jacobian determinant and prescribed curl vector.
The problem for diffeomorphism with prescribed Jacobian determinant has been solved by the deformation method \cite{Liao2004, Chen2016}. But the prescribed Jacobian determinant alone cannot uniquely determine a diffeomorphism. In \cite{Chen2016,Chen2015}, an innovative variational method is proposed for numerical construction of a diffeomorphism with prescribed Jacobian determinant and prescribed curl vector.\\
%this paper we will present theoretical evidence that prescribed Jacobian and curl together uniquely determine the diffeomorphism.
This new method enables us to define a new approach to robustly averaging given diffeomorphisms so that the average is guaranteed to be a diffeomorphism. In fact, this research is motivated by building of medical image atlases. An atlas is a model image of an organ constructed by "averaging" images of a group of healthy individuals or patients \cite{Evans2012, Mandal2012}.  Atlases can be used to detect and monitor health problems and diseases such as Alzheimer's disease.  Constructing a correct atlas is a very challenging task. This is mainly due to the rich variability of anatomy in individuals of the population. In particular, both the building and use of an atlas are greatly dependent on non-rigid image registration techniques which bring the images to a common coordinate system by nonlinear transformations. Early methods \cite{Thompson2002} of building an atlas from a set of images consist of two steps: (1) registering each image to an initial template; (2) averaging the warped (resampled) images.\\
In order to reduce the blurs of the atlas due to registration errors, we adopt the idea of \cite{Vaillant2004,Twining2008} and \cite{Asigny2006} that the registration transformations are averaged first and then the template image is re-sampled in the averaged transformation. Thus we build an atlas in three steps: (1) constructing an initial template or simply selecting one of the images as a template; and then registering all images to the template by diffeomorphisms; (2) averaging the registration diffeomorphisms; and (3) re-sampling the template image on the averaged diffeomorphism. This paper is mainly focused on the second step: averaging the diffeomorphisms.\\
Since the goal of image atlases is to quantify variability and to detect abnormality in individuals, brain image atlases must be correctly built. But current methods of building and using atlases are not the best possible in terms of their theoretical foundation, reliability, and computational efficiency. A thorough review of available literatures indicates that there is no agreement in medical imaging community on how the ※average§ of diffeomorphisms should be defined and computed. The widely used Euclidean average may not have the averaged size and may even fail to be a diffeomorphism. This is illustrated by graphics below. Other methods proposed are restricted to the particular diffeomorphisms that are generated by certain registration methods. Moreover, these methods again do not guarantee that the ※average§ have the averaged size or averaged local rotation. From a mathematical point of view, the main problem is how to generate diffeomorphisms with prescribed Jacobian determinant and prescribed curl vector. In this paper, we propose an innovative approach to averaging general diffeomorphisms, including those arising from brain imaging atlases. In Section 2, a new definition of the average is defined. In Section 3, the variational method for construction of diffeomorphisms based on the Jacobian determinant and the curl vector is briefly reviewed. In Section 4, numerical examples are presented.

\section{Average of Diffeomorphisms}
\subsection{The Euclidean Average}
The commonly used Euclidean average of a set of transformations is to simply average the Cartesian coordinates of the points specified by the transformations.\\
We now examine what can go wrong with the Euclidean average and convince the readers that the correct average must have the averaged size and the averaged local rotation. To illustrate this point, we use a grid to approximate a transformation. Each grid node in a uniform grid on $\Omega = [0, 1]\times[0, 1]$ is sent to a point in $\Omega$. We keep the same connectivity between the adjacent nodes and get a new grid in $\Omega$, which is a good approximation if the grid spacing is very small. Let $\Phi_1$ be the identity transformation from $\Omega$ to $\Omega$; let $\Phi_2$ be a diffeomorphism from $\Omega$ to $\Omega$. A small cell with corners $\{A, B, C, D\}$ is sent by $\Phi_1$ to the same cell in $\Omega$; We now consider three examples of $\Phi_2$. First, suppose $\Phi_2$ is locally a translation followed by a rotation counterclockwise in the angle of $0^\circ$, which sends the cell $\{A, B, C, D\}$ to the cell $\{A', B', C', D'\}$ shown in Fig. 1(a). The Euclidean average of the four pairs of corners are denoted by $\{\overline{A},\overline{B},\overline{C},\overline{D}\}$. Note that the cell formed by $\{\overline{A},\overline{B},\overline{C},\overline{D}\}$ is a meaningful average of the two cells $\{A, B, C, D\}$ and $\{A', B', C', D'\}$. But when the rotation angle increases from $0^\circ$ to $90^\circ$ and to $180^\circ$, the resulting Euclidean average is deteriorating as shown in Fig. 1(b) and (c). Suppose $\Phi_2$ is locally the same translation followed by a rotation counterclockwise in the angle of $90^\circ$ which sends the cell $\{A, B, C, D\}$ to the cell $\{A', B', C', D'\}$ shown in Fig. 1(b). The Euclidean average $\{\overline{A},\overline{B},\overline{C},\overline{D}\}$ is too small to be a meaningful average of the two cells $\{A, B, C, D\}$ and $\{A', B', C', D'\}$. Suppose now $\Phi_2$ is locally the same translation followed by a rotation counterclockwise in the angle of $180^\circ$, which sends the same grid cell to the cell $\{A', B', C', D'\}$ shown in the Fig. 1(c). Then the Euclidean average, cell $\{\overline{A},\overline{B},\overline{C},\overline{D}\}$ shrinks to a point P. This means the ※cell§ degenerates to a point and the Jacobian determinant becomes zero! These figures show clearly that local rotations play a crucial role in the study of shapes. The role of local rotation will also be demonstrated in Figure 2 below.
\begin{figure}
  \begin{center}
    \begin{tikzpicture}
    %a
    \coordinate [label=below left:$A$] (A) at (0,0);
    \coordinate [label=below right:$B$] (B) at (2,0);
    \coordinate [label=above left:$C$] (C) at (2,2);
    \coordinate [label=above left:$D$] (D) at (0,2);
    \coordinate (v) at (0.5,3);
    \coordinate [label=below right:$A'$] (A1) at ($(A)+(v)$);
    \coordinate [label=right:$B'$] (B1) at ($(B)+(v)$);
    \coordinate [label=above right:$C'$] (C1) at ($(C)+(v)$);
    \coordinate [label=above left:$D'$] (D1) at ($(D)+(v)$);
      \draw [black,very thick] (A)--(B)--(C)--(D)--(A);
      \draw [black,very thick] (A1)--(B1)--(C1)--(D1)--cycle;
      \draw[red,dashed] (A)--coordinate[midway](Ab)(A1);
      \draw[red,dashed] (B)--coordinate[midway](Bb)(B1);
      \draw[red,dashed] (C)--coordinate[midway](Cb)(C1);
      \draw[red,dashed] (D)--coordinate[midway](Db)(D1);
      \draw[blue,very thick] (Ab) node[below right]{$\overline{A}$} --(Bb) node[below right]{$\overline{B}$}--(Cb)node[above left]{$\overline{C}$}--(Db) node[above left]{$\overline{D}$}--cycle;
      \node [below,align=center] at (1,-1) {(a) Rotation=$0^\circ$};
    %b
    \coordinate (u) at (4,0);
    \coordinate [label=below left:$A$] (A) at ($(A)+(u)$);
    \coordinate [label=below right:$B$] (B) at ($(B)+(u)$);
    \coordinate [label=above left:$C$] (C) at ($(C)+(u)$);
    \coordinate [label=above left:$D$] (D) at ($(D)+(u)$);
    \coordinate (v) at (0.5,3);
    \coordinate [label=right:$A'$] (A1) at ($(B)+(v)$);
    \coordinate [label=above right:$B'$] (B1) at ($(C)+(v)$);
    \coordinate [label=above left:$C'$] (C1) at ($(D)+(v)$);
    \coordinate [label=above left:$D'$] (D1) at ($(A)+(v)$);
      \draw [black,very thick] (A)--(B)--(C)--(D)--(A);
      \draw [black,very thick] (A1)--(B1)--(C1)--(D1)--cycle;
      \draw[red,dashed] (A)--coordinate[midway](Ab)(A1);
      \draw[red,dashed] (B)--coordinate[midway](Bb)(B1);
      \draw[red,dashed] (C)--coordinate[midway](Cb)(C1);
      \draw[red,dashed] (D)--coordinate[midway](Db)(D1);
      \draw[blue,very thick] (Ab) node[below right]{$\overline{A}$} --(Bb) node[below right]{$\overline{B}$}--(Cb)node[above left]{$\overline{C}$}--(Db) node[above left]{$\overline{D}$}--cycle;
      \node [below,align=center] at ($(1,-1)+(u)$) {(b) Rotation=$90^\circ$};
    %c
    \coordinate (u) at (4,0);
    \coordinate [label=below left:$A$] (A) at ($(A)+(u)$);
    \coordinate [label=below right:$B$] (B) at ($(B)+(u)$);
    \coordinate [label=above left:$C$] (C) at ($(C)+(u)$);
    \coordinate [label=above left:$D$] (D) at ($(D)+(u)$);
    \coordinate (v) at (0.5,3);
    \coordinate [label=above right:$A'$] (A1) at ($(C)+(v)$);
    \coordinate [label=above left:$B'$] (B1) at ($(D)+(v)$);
    \coordinate [label=below left:$C'$] (C1) at ($(A)+(v)$);
    \coordinate [label=below right:$D'$] (D1) at ($(B)+(v)$);
      \draw [black,very thick] (A)--(B)--(C)--(D)--(A);
      \draw [black,very thick] (A1)--(B1)--(C1)--(D1)--cycle;
      \draw[red,dashed] (A)--coordinate[midway](Ab)(A1);
      \draw[red,dashed] (B)--coordinate[midway](Bb)(B1);
      \draw[red,dashed] (C)--coordinate[midway](Cb)(C1);
      \draw[red,dashed] (D)--coordinate[midway](Db)(D1);
      \draw[blue, very thick] (Ab) node[above right]{$P$};
      \node [below,align=center] at ($(1,-1)+2*(u)$) {(c) Rotation=$180^\circ$};
    \end{tikzpicture}
  \caption{Euclidean average of transformation $\Phi_1$ and different $\Phi_2$.}
  \end{center}
\end{figure}
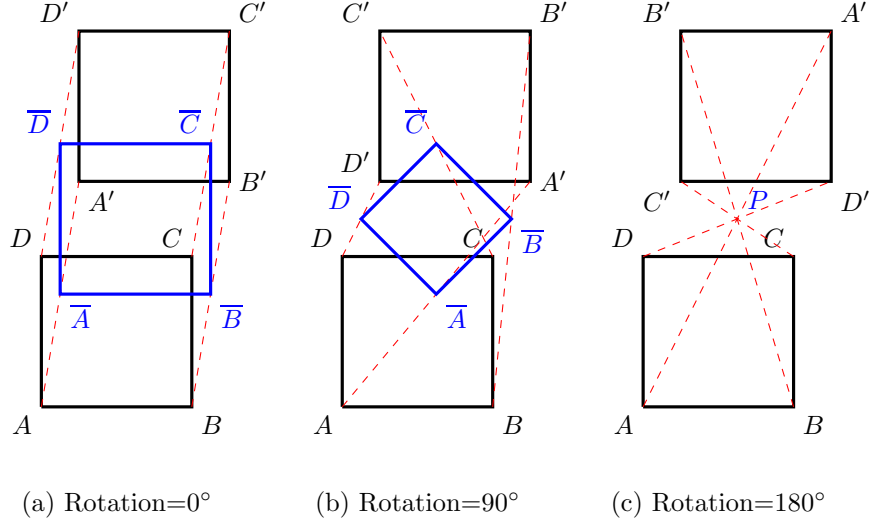
%\begin{figure}[H]
%\begin{center}
%\subfigure[Rotation=$0^\circ$]{\includegraphics[width=0.30\textwidth]{Ex1_a.jpg}}
%\subfigure[Rotation=$90^\circ$]{\includegraphics[width=0.30\textwidth]{Ex1_b.jpg}}
%\subfigure[Rotation=$180^\circ$]{\includegraphics[width=0.38\textwidth]{Ex1_c.jpg}}
%\caption{Euclidean average of transformation $\Phi_1$ and different $\Phi_2$}
%\end{center}
%\end{figure}

\subsection{Some Non-Euclidean Approaches}
There are currently two non-Euclidean approaches to averaging of diffeomorphisms arising from nonlinear image registration. Miller\'s group at Johns Hopkins University has been a main contributor in this research area. Their approach (see \cite{Vaillant2004}) is based on their image registration method called the Large Deformation Diffeomorphic Metric Matching (LDDMM) \cite{Beg2005} by time-varying velocity fields. The theoretical foundation of their approach is that a certain quantity called the ※moment§ is time invariant in their method of image registration. Thus the moments of diffeomorphisms at $t=0$ determine the diffeomorphisms generated at any time $t>0$. They perform statistics such as determining the average on the initial moments. Then, the average of these diffeomorphisms is generated from this averaged moments by their method of image registration.
There are two main issues with this approach: (1) It only treats a special type of diffeomorphisms that are generated by their specific methods of image registration; but these diffeomorphisms are rather restrictive as can be seen in their experiments in \cite{Beg2005} (see Examples below); (2) The numerical implementation of the approach is sensitive since the moment is in fact the inverse of the Green＊s function of a linear differential operator (i.e. the Laplace operator $\Delta+0.1\cdot Identity$), and therefore exists only in weak sense. As a consequence, in \cite{Vaillant2004} it was only formulated and tested for diffeomorphisms constructed by matching a few dozen landmark points in one of the two images ($256\times256\times64$), with the same number of landmark points in the other. Seven years have passed and there has been no work published for more general diffeomorphisms that are generated based on image intensity.\\
\cite{Twining2008} described in detail the mathematical foundation of the Large Deformation Diffeomorphic Metric Matching (LDDMM) and proposed a variant to the approach in \cite{Vaillant2004}. This paper considers the problem of defining an average of diffeomorphisms, motivated primarily by image registration problem, where diffeomorphisms are used to align images of large deformation. Constructing an average on the diffeomorphism group will enable the quantitative analysis of these diffeomorphisms to discover the normal and abnormal variation of structures in a population. Based on splines, they construct an average for particular choices of boundary conditions on the space on which the diffeomorphism acts, and for a particular class of metrics on the diffeomorphism group, which define a class of diffeomorphic interpolating splines. The geodesic equation is computed for this class of metrics, and they show how it can be solved in the spline representation. Furthermore, they demonstrate that the spline representation generates submanifolds of the diffeomorphism group. Instead of matching land mark points as in \cite{Vaillant2004}, their approach is based on matching spline control nodes in the images. It determines the usual Euclidean average of control nodes of a set of diffeomorphisms, and then generates the ※average§ of these diffeomorphisms by spline interpolation. Explicit computational examples are included, showing how this average can be constructed in practice, and that the use of the geodesic distance allows better classification of variation than those obtained using just a Euclidean metric on the space of diffeomorphisms. This method is an variant of the method in \cite{Vaillant2004} and is only suitable to diffeomorphisms generated by the Large Deformation Diffeomorphic Metric Matching (LDDMM).\\
In \cite{Avants2011}, another variant to the method in \cite{Vaillant2004} is proposed. It proposed a method of atalas building that is based on their version of the Large Deformation Diffeomorphic Metric Matching (LDDMM), called the symmetric normalization (SyN). It is included as part of an open source software package called the Advanced Neuroimaging Tools (ANTs) developed at University of Pennsylvania. The image registration method SyN received a top ranking in \cite{Klein2009} and a groupwise version called SyGN is used to build an atlas from a group of images. As commented before, as an variant, SyN uses a smoothing term to regularize the optimization of a similarity measure. The smoothing term is based on a linear differential operator $L= a \cdot\Delta + b\cdot identity$, which inevitably altered the optimization problem: a trade-off between the similarity and a smoothing term is optimized. Instead, we should optimize the similarity measure in the space of all diffeomorphisms, not on a subspace defined by a linear differential operator $L$. Thus, there is room for significant improvement in registration accuracy. Also, this atlas building method works with diffeomorphisms arising from their groupwise symmetric normalization (SyGN) only.\\
In \cite{Asigny2006}, a different approach is proposed which in theory is suitable only for small deformation in images, i.e. it works in theory only for diffeomorphisms that are near the identity transformation. The approach, called the Log-Euclidean is also based on generating diffeomorphisms from a time-independent velocity field. The diffeomorphisms this approach treats are even more restrictive compared to the previous approach. The approach is theoretically valid only for diffeomorphisms near the identity. Nonetheless, their numerical algorithms appear to be robust and fast, and numerical experiments can still go through for diffeomorphisms that are not near the identity; but, the author indicated that the results obtained by the Log-Euclidean are not very different from that of the Euclidean average (which may not be diffeomorphic).

\subsection{Proposed Approach to Averaging Diffeomorphisms}
Let $\Omega = [0,1]\times[0,1]$ in $R^2$ or $\Omega= [0,1]\times[0,1]\times[0,1]$ in $R^3$. Suppose there are diffeomorphisms $\Phi_i, i = 1, 2\ldots K$, from $\Omega$ into itself such that $\Phi_i = identity$ on $\partial\Omega$ (other boundary conditions are possible). Our concept of average is based on Jacobian determinants and the curl vector field. The former controls the size of volume elements; the latter controls the local rotation. We can show both theoretically and computationally that together these two quantities under proper boundary conditions uniquely determine a transformation \cite{Liao2009, Chen2016}. \\
To motivate our definition of the average, let us revisit Fig.1(a), (b), (c), and ask ourselves what is the correct average of the cells $\{A, B, C, D\}$ and $\{A', B', C', D'\}$ in each of the three cases.
For Fig. 1(a), we should simply take the cell $\{\overline{A},\overline{B},\overline{C},\overline{D}\}$ as the average since it has the correct size and correct orientation. This expected mean is denoted by $\{A^*, B^*, C^*, D^*\}$ in Fig. 2(a). For Fig. 1(b), the cell $\{\overline{A},\overline{B},\overline{C},\overline{D}\}$  has the correct orientation since it forms $45^\circ$ with both cell $\{A, B, C, D\}$ and cell $\{A', B', C', D'\}$. But its size is too small. Since cell $\{A, B, C, D\}$ and cell $\{A', B', C', D'\}$ have the same size, we expect their average to have the same size also. The expected average is denoted by $\{A^*, B^*, C^*, D^*\}$ in Fig. 2(b). For Fig. 2(c), the cell $\{\overline{A},\overline{B},\overline{C},\overline{D}\}$ is reduced to the point P. Since the relative rotation is $180^\circ$, the expected average should form a $90^\circ$ angle with both cell $\{A, B, C, D\}$ and cell $\{A', B', C', D'\}$. Its size should also be the same as cell $\{A, B, C, D\}$ and cell $\{A', B', C', D'\}$. The cell denoted in Fig. 2(c) by $\{A^*, B^*, C^*, D^*\}$ is the expected average.\\
\begin{figure}
  \begin{center}
    \begin{tikzpicture}
    %a
    \coordinate [label=below left:$A$] (A) at (0,0);
    \coordinate [label=below right:$B$] (B) at (2,0);
    \coordinate [label=above left:$C$] (C) at (2,2);
    \coordinate [label=above left:$D$] (D) at (0,2);
    \coordinate (v) at (0.5,3);
    \coordinate [label=below right:$A'$] (A1) at ($(A)+(v)$);
    \coordinate [label=right:$B'$] (B1) at ($(B)+(v)$);
    \coordinate [label=above right:$C'$] (C1) at ($(C)+(v)$);
    \coordinate [label=above left:$D'$] (D1) at ($(D)+(v)$);
      \draw [black,very thick] (A)--(B)--(C)--(D)--(A);
      \draw [black,very thick] (A1)--(B1)--(C1)--(D1)--cycle;
      \draw[red,dashed] (A)--coordinate[midway](Ab)(A1);
      \draw[red,dashed] (B)--coordinate[midway](Bb)(B1);
      \draw[red,dashed] (C)--coordinate[midway](Cb)(C1);
      \draw[red,dashed] (D)--coordinate[midway](Db)(D1);
      \draw[green,very thick] (Ab) node[below right]{$A^*$} --(Bb) node[below right]{$B^*$}--(Cb)node[above left]{$C^*$}--(Db) node[above left]{$D^*$}--cycle;
      \node [below,align=center] at (1,-1) {(a) Rotation=$0^\circ$};
    %b
    \coordinate (u) at (4,0);
    \coordinate [label=below left:$A$] (A) at ($(A)+(u)$);
    \coordinate [label=below right:$B$] (B) at ($(B)+(u)$);
    \coordinate [label=above left:$C$] (C) at ($(C)+(u)$);
    \coordinate [label=above left:$D$] (D) at ($(D)+(u)$);
    \coordinate (v) at (0.5,3);
    \coordinate [label=right:$A'$] (A1) at ($(B)+(v)$);
    \coordinate [label=above right:$B'$] (B1) at ($(C)+(v)$);
    \coordinate [label=above left:$C'$] (C1) at ($(D)+(v)$);
    \coordinate [label=above left:$D'$] (D1) at ($(A)+(v)$);
      \draw [black,very thick] (A)--(B)--(C)--(D)--(A);
      \draw [black,very thick] (A1)--(B1)--(C1)--(D1)--cycle;
      \draw[red,dashed] (A)--coordinate[midway](Ab)(A1);
      \draw[red,dashed] (B)--coordinate[midway](Bb)(B1);
      \draw[red,dashed] (C)--coordinate[midway](Cb)(C1);
      \draw[red,dashed] (D)--coordinate[midway](Db)(D1);
      \draw[blue,very thick] (Ab) node[below right]{$\overline{A}$} --(Bb) node[below right]{$\overline{B}$}--(Cb)node[above left]{$\overline{C}$}--(Db) node[above left]{$\overline{D}$}--cycle;
      \draw[green,very thick] (5.25,1.0858) node[below]{$A^*$} --(6.6642,2.5) node[right]{$B^*$}--(5.25,3.9142)node[above]{$C^*$}--(3.8358,2.5) node[left]{$D^*$}--cycle;
      \node [below,align=center] at ($(1,-1)+(u)$) {(b) Rotation=$90^\circ$};
    %c
    \coordinate (u) at (4,0);
    \coordinate [label=below left:$A$] (A) at ($(A)+(u)$);
    \coordinate [label=below right:$B$] (B) at ($(B)+(u)$);
    \coordinate [label=above left:$C$] (C) at ($(C)+(u)$);
    \coordinate [label=above left:$D$] (D) at ($(D)+(u)$);
    \coordinate (v) at (0.5,3);
    \coordinate [label=above right:$A'$] (A1) at ($(C)+(v)$);
    \coordinate [label=above left:$B'$] (B1) at ($(D)+(v)$);
    \coordinate [label=below left:$C'$] (C1) at ($(A)+(v)$);
    \coordinate [label=below right:$D'$] (D1) at ($(B)+(v)$);
      \draw [black,very thick] (A)--(B)--(C)--(D)--(A);
      \draw [black,very thick] (A1)--(B1)--(C1)--(D1)--cycle;
      \draw[red,dashed] (A)--coordinate[midway](Ab)(A1);
      \draw[red,dashed] (B)--coordinate[midway](Bb)(B1);
      \draw[red,dashed] (C)--coordinate[midway](Cb)(C1);
      \draw[red,dashed] (D)--coordinate[midway](Db)(D1);
      \draw[blue, very thick] (Ab) node[above right]{$P$};
      \path (A)--coordinate[midway](Ab)(C1);
      \path (B)--coordinate[midway](Bb)(D1);
      \path (C)--coordinate[midway](Cb)(A1);
      \path (D)--coordinate[midway](Db)(B1);
      \draw[green,very thick] (Ab) node[below]{$D^*$} --(Bb) node[below right]{$A^*$}--(Cb)node[above]{$B^*$}--(Db) node[above left]{$C^*$}--cycle;
      \node [below,align=center] at ($(1,-1)+2*(u)$) {(c) Rotation=$180^\circ$};
    \end{tikzpicture}
  \caption{Expected average of transformation $\Phi_1$ and different $\Phi_2$.}
  \end{center}
\end{figure}
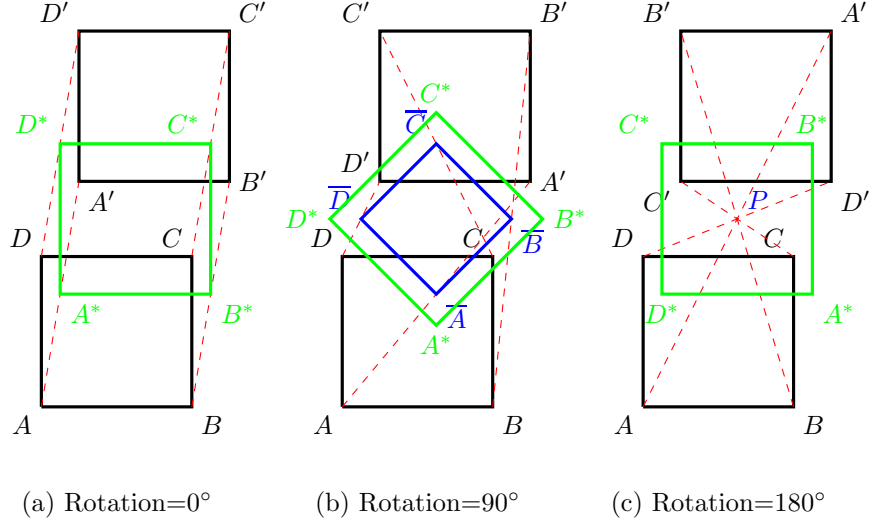
%\begin{figure}[H]
%\begin{center}
%\subfigure[Rotation=$0^\circ$]{\includegraphics[width=0.28\textwidth]{Ex2_a.jpg}}
%\subfigure[Rotation=$90^\circ$]{\includegraphics[width=0.35\textwidth]{Ex2_b.jpg}}
%\subfigure[Rotation=$180^\circ$]{\includegraphics[width=0.30\textwidth]{Ex2_c.jpg}}
%\caption{Expected average of transformation $\Phi_1$ and different $\Phi_2$}
%\end{center}
%\end{figure}
Next, we introduce the definition of the average and describe our approach to computing it by differential equations.\\
\textbf{Proposed Definition of the Average of Diffeomorphisms:}\\
The average of given diffeomorphisms $\Phi_i, i = 1, 2\ldots K$, is defined as the diffeomorphism $\Phi$ which has the following properties:
\begin{enumerate}
  \item $J(\Phi)=\sum w_i J(\Phi_i)$, where $J$ is the Jacobian determinant of a diffeomorphism,
  \item $curl (\Phi) =\sum w_i curl(\Phi_i)$,  where $curl$ is the curl vector field of a diffeomorphism;
\end{enumerate}
and $w_i$'s are positive weights whose sum is $1$. (For instance, we may simply take all $w_i = 1/K$; or we can define $w_i$ as proportional to the distance between $\Phi_i(x)$ and the Euclidean mean $\sum \Phi_i/K$).\\
Thus, mathematically, the problem of determining the average of diffeomorphisms $\Phi_i, i = 1, 2\ldots K$, is to construct a diffeomorphism with prescribed Jacobian determinant $f=\sum w_i J(\Phi_i)$ and the prescribed curl vector field $g =\sum w_i curl(\Phi_i)$ under proper boundary conditions.\\
In next section, we present a variational method which computes a diffeomorphism $\Phi$ with prescribed Jacobian determinant and curl robustly and efficiently.

\section{A Variational Method}
The most technical step is to construct a diffeomorphism with prescribed Jacobian determinant and the prescribed curl vector. This problem is formulated as follows:\\
Given $f_0>0$ and $\bm{g}_0$ on $\Omega$, minimize the energy functional
\[
E(\bm{\Phi},f_0,\bm{g}_0)=\frac 1 2\int_\Omega[(J(\bm{\Phi}(x))-f_0(x))^2+(curl(\bm{\Phi}(x))-\bm{g}_0(x))^2]dx
\]
subject to the constraints:
\[
\bm{\Phi}(x)=\bm{\Phi_0}(x)+\bm{u}(x),
\]
where $\bm{u}$ satisfies:
\[
\left\{
\begin{array}{ll}
\textrm{div} \bm{u}&=f\\
\textrm{curl} \bm{u}&=\bm{g} \qquad in \ \Omega\\
\bm{u}&=0 \qquad on \ \partial\Omega.
\end{array}
\right.
\]
In the numerical optimization, the \textrm{div} and \textrm{curl} equation are replaced by
\[
\Delta \bm{u}=(f_1,f_2,f_3)=\bm{F}.
\]
Thus, the diffeomorphism $\Phi$ is constructed by optimizing $E(\bm{\Phi},f_0,\bm{g}_0)$ with respect to the control function $\bm{F}=(f_1,f_2,f_3)$. Note: the Jacobian determinant $J(\bm{\Phi})$ and the curl vector $curl(\bm{\Phi})$ are equal to $f_0$ and $\bm{g}_0$ in the $L_2-$ norm.\\
In \cite{Chen2016, Chen2015}, the gradient $\frac{\partial E}{\partial \bm{F}}$ of $E(\bm{\Phi},f_0,\bm{g}_0)$ is derived. \\
%Indeed, $\frac{\partial E}{\partial \bm{F}}=\bm{G}=(g_1,g_2,g_3)$ where $g_i$ satisfies\\
An optimization algorithm is implemented based on the method of gradient descent. In the following example, the coordinates of the grid in Fig 3. are used to represent the diffeomorphism, and to calculate the Jacobian determinant $f_0$ and the curl vector $\bm{g}_0$. Our algorithm generated a grid that is almost identical to the original grid in Fig 4,5.
\begin{figure}[H]
  \begin{center}
    \begin{tikzpicture}
    %\draw[help lines]  (0,0) grid (6,2);
      \node at(0,0) {\includegraphics[width=0.31\textwidth]{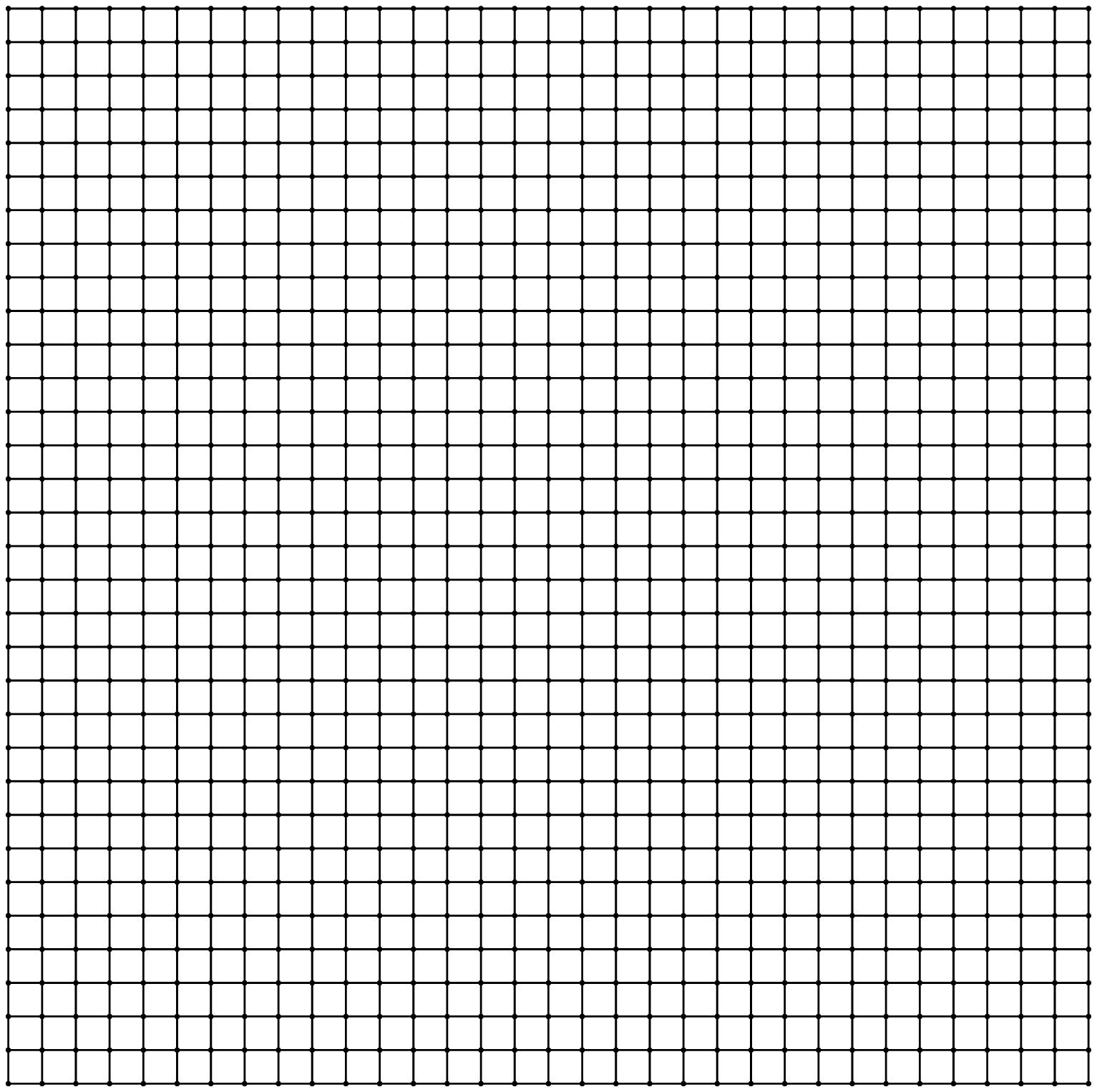}};
      \node at(6,0) {\includegraphics[width=0.32\textwidth]{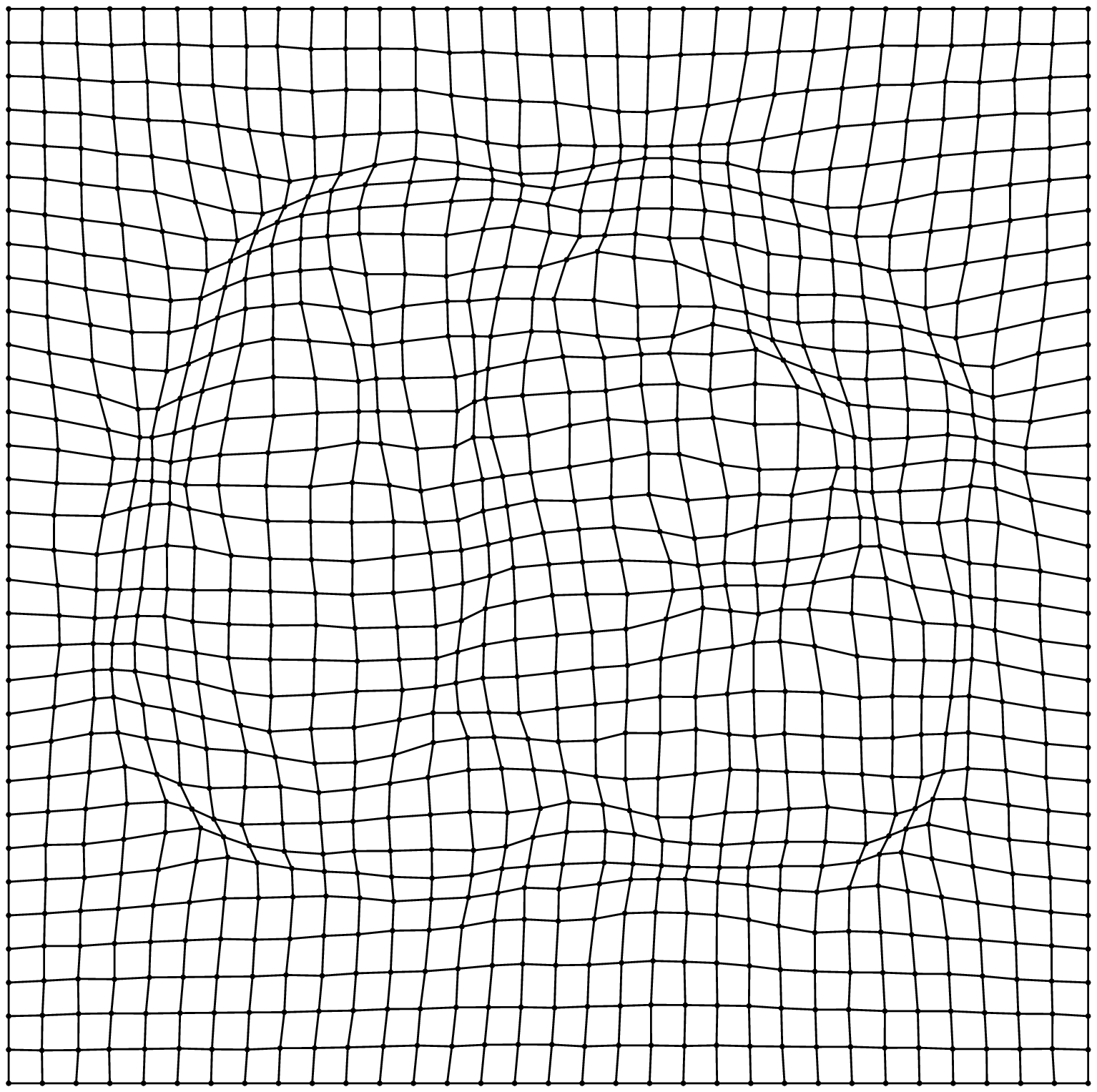}};
      \draw [->,blue, very thick] (1.7,0) to [out=30,in=150] node[midway,above]{$\Phi_0$} (4.2,0);
    \end{tikzpicture}

  \caption{Test transformation $\Phi_0$}
  \end{center}
\end{figure}

%\begin{figure}[H]
%\begin{center}
%%\subfigure[$\Phi_0$]{\includegraphics[width=0.32\textwidth]{phi_0.eps}}
%
%\subfigure[Reconstructed $\Phi_0^{(1)}$, having $E(\bm{\Phi},f_0,\bm{g}_0)$ decreases by 90\% ]{\includegraphics[width=0.32\textwidth]{phi_0_rec_01_1.eps}}
%\subfigure[enlarged view of left blue rectangle]{\includegraphics[width=0.32\textwidth]{phi_0_rec_01_2.eps}}
%\subfigure[enlarged view of right blue rectangle]{\includegraphics[width=0.32\textwidth]{phi_0_rec_01_3.eps}}
%
%\subfigure[Reconstructed $\Phi_0^{(2)}$, having $E(\bm{\Phi},f_0,\bm{g}_0)$ decreases by 99\% ]{\includegraphics[width=0.32\textwidth]{phi_0_rec_001_1.eps}}
%\subfigure[enlarged view of left blue rectangle]{\includegraphics[width=0.32\textwidth]{phi_0_rec_001_2.eps}}
%\subfigure[enlarged view of right blue rectangle]{\includegraphics[width=0.32\textwidth]{phi_0_rec_001_3.eps}}
%\caption{Reconstruction of a transformation grid from its cell size and local rotation. \footnotesize{The black star dots $*$ represent $\bm{\Phi}_0$, and red dots $\cdot$ represent constructed $\bm{\Phi}_0^{(1,2)}$}}
%\end{center}
%\end{figure}
\begin{figure}[H]
\begin{center}
\begin{tikzpicture}
%\draw [help lines] (0,0) grid (8,5);
\node[inner sep=0pt] (full) at (0,0)
    {\includegraphics[width=.35\textwidth]{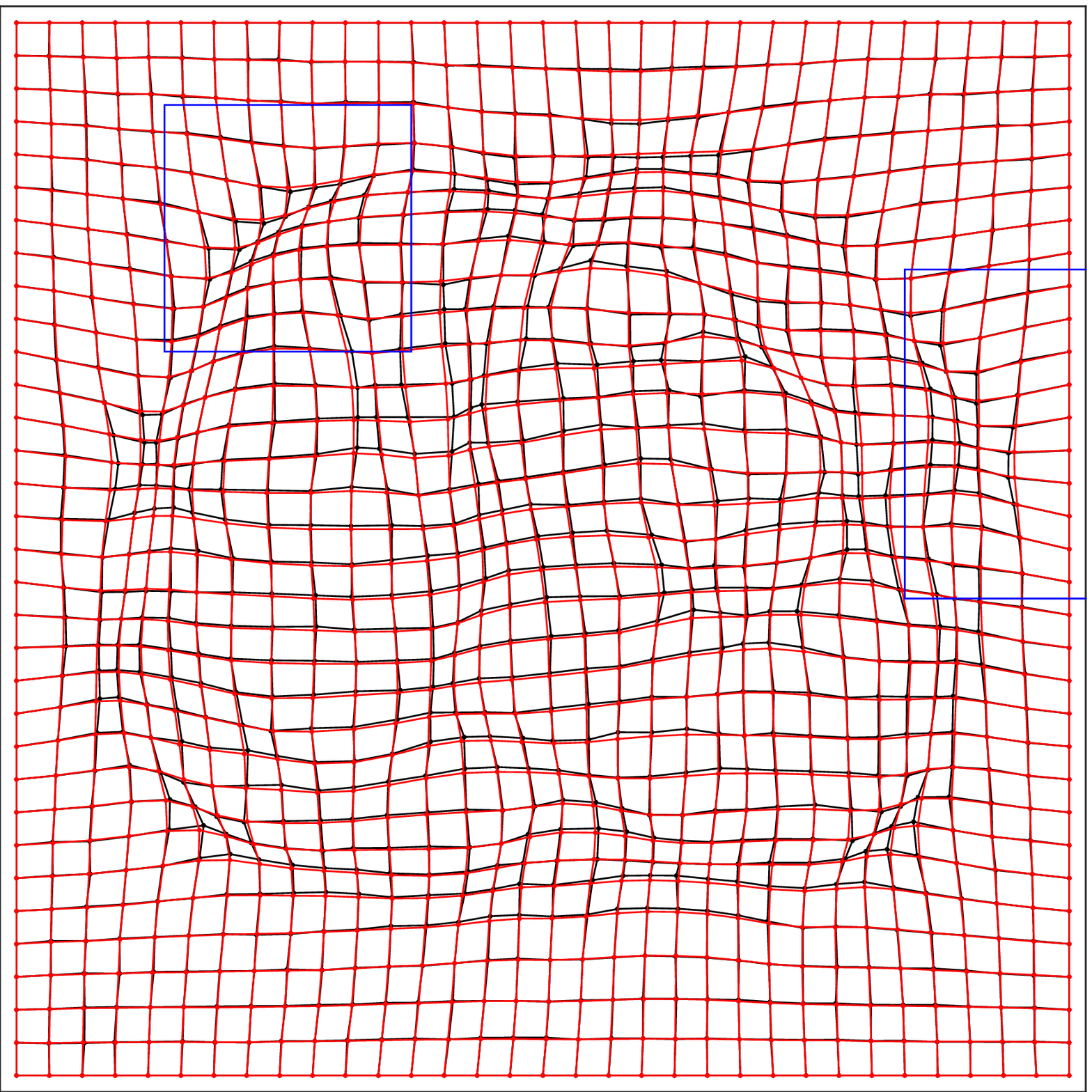}};
\node [below,align=center] at (0,-2) {(a) Reconstructed $\Phi_0^{(1)}$, having\\ $E(\bm{\Phi},f_0,\bm{g}_0)$ decreases by 90\%};
\node[inner sep=0pt] (part1) at (5,3)
    {\includegraphics[width=.35\textwidth]{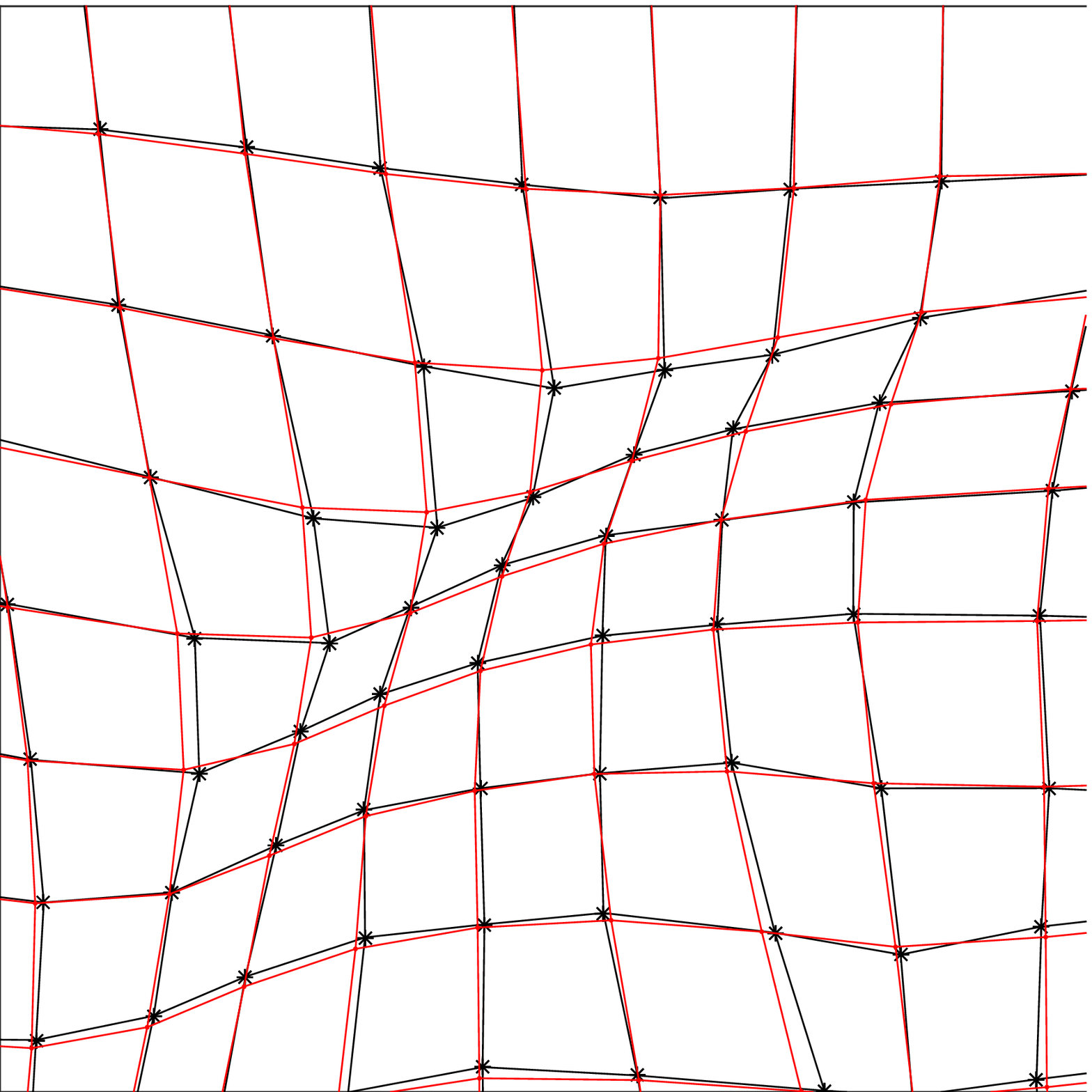}};
\node [below,align=center] at (5,1) {(b) Enlarged view of the left\\ blue rectangle};
\node[inner sep=0pt] (part1) at (5,-2)
    {\includegraphics[width=.35\textwidth]{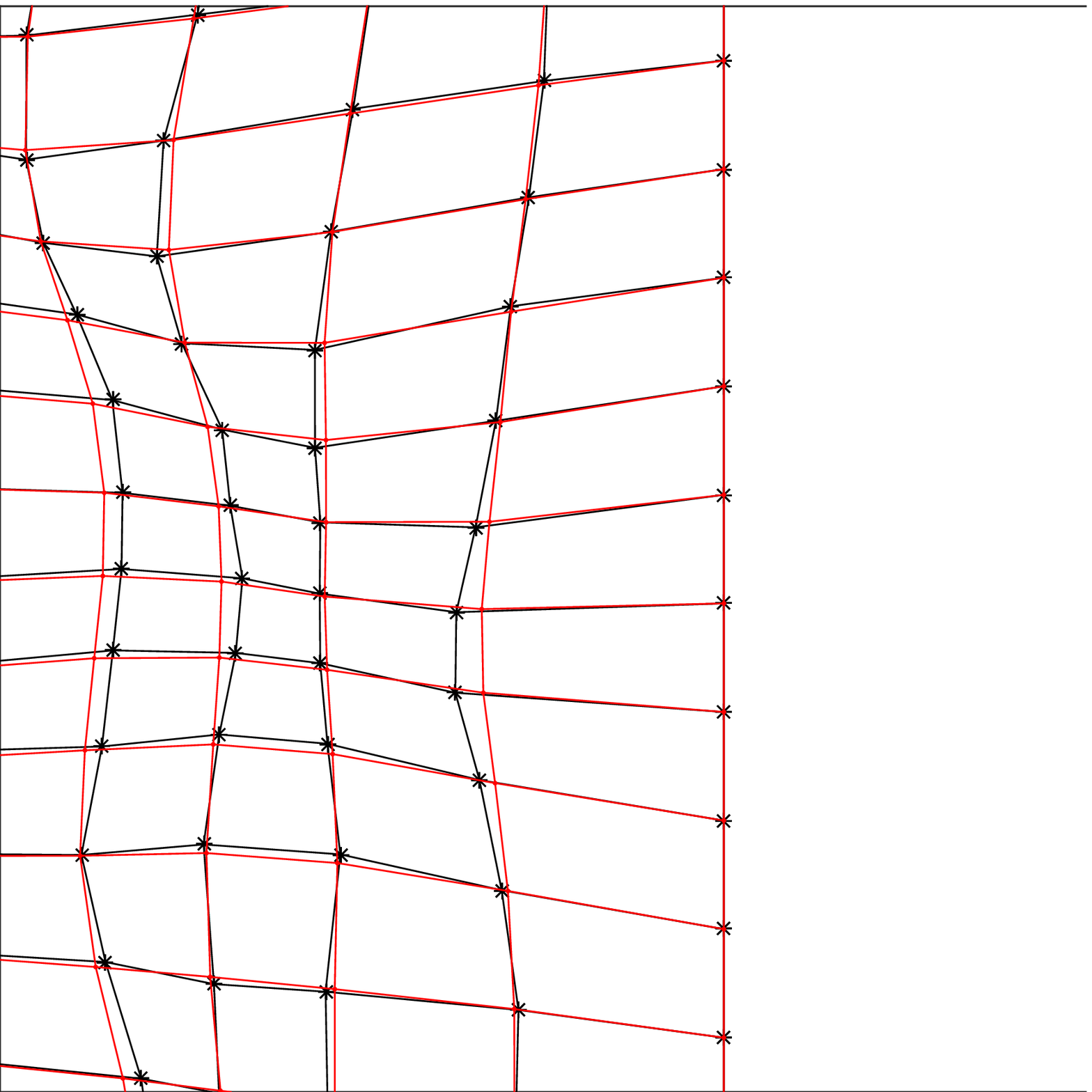}};
\node [below,align=center] at (5,-4) {(c) Enlarged view of the right\\ blue rectangle};
\draw [->,blue] (-1,1.8) to [out=90,in=180] (2.9,3);
\draw [->,blue] (2.1,0.5) to [out=0,in=180] (2.9,-2);
\end{tikzpicture}
\caption{Reconstruction of a transformation grid from its cell size and local rotation. \footnotesize{The black star dots $*$ represent $\bm{\Phi}_0$, and red dots $\cdot$ represent constructed $\bm{\Phi}_0^{(1)}$}, $65\times65$ grid size, 0.62s by a Matlab program on a general laptop.}
\end{center}
\end{figure}

\begin{figure}[H]
\begin{center}
\begin{tikzpicture}
%\draw [help lines] (0,0) grid (8,5);
\node[inner sep=0pt] (full) at (0,0)
    {\includegraphics[width=.35\textwidth]{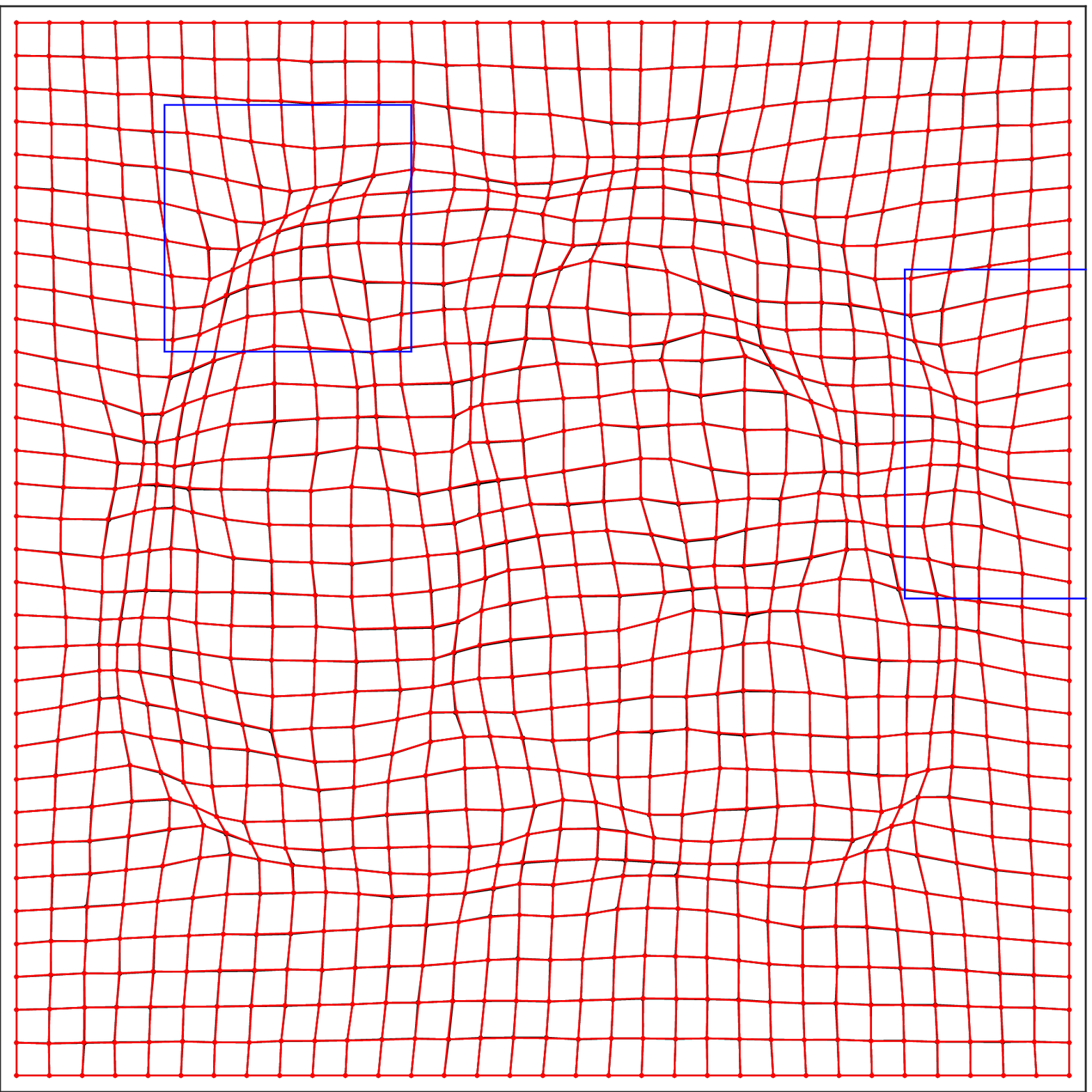}};
\node [below,align=center] at (0,-2) {(a) Reconstructed $\Phi_0^{(2)}$, having\\ $E(\bm{\Phi},f_0,\bm{g}_0)$ decreases by 99\%};
\node[inner sep=0pt] (part1) at (5,3)
    {\includegraphics[width=.35\textwidth]{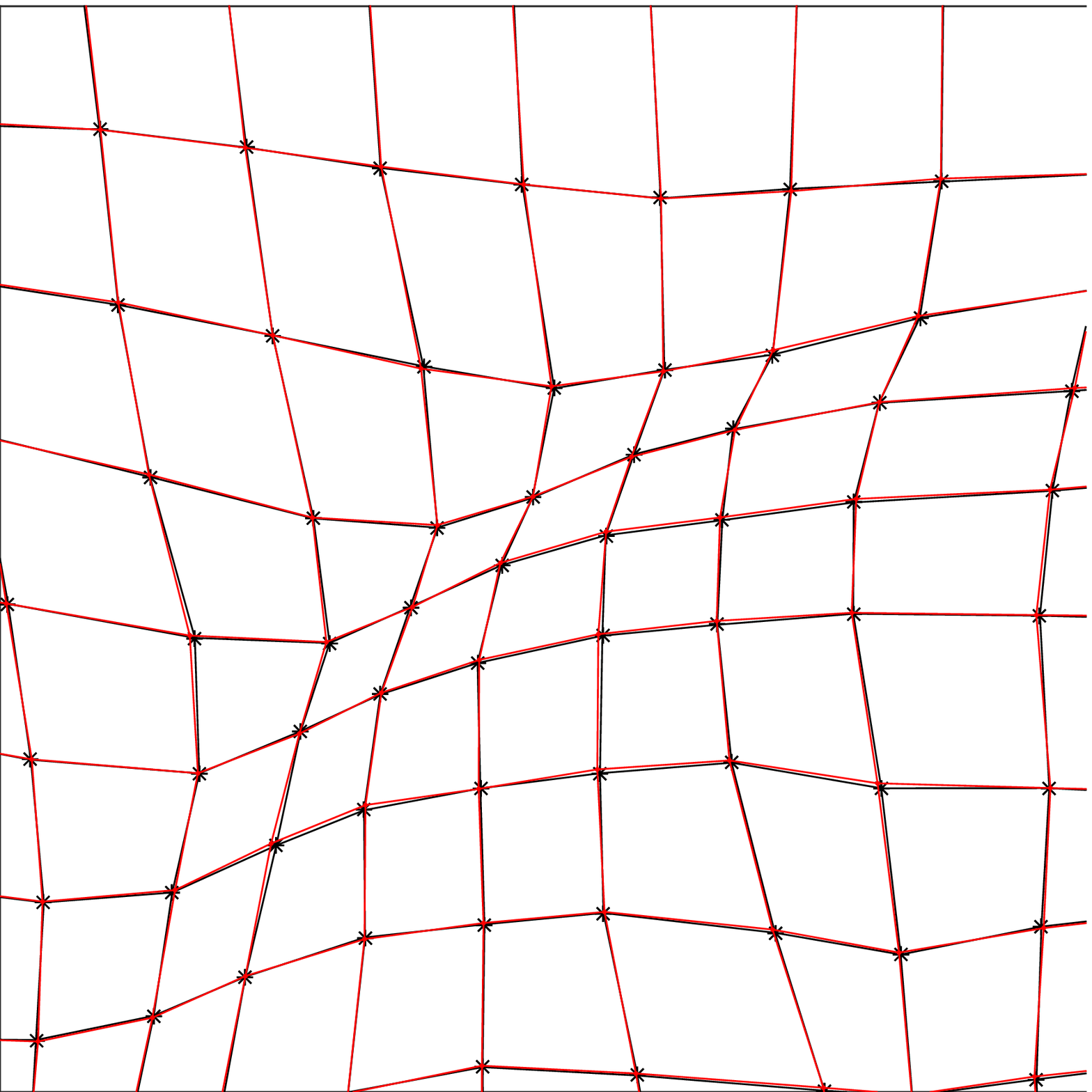}};
\node [below,align=center] at (5,1) {(b) Enlarged view of the left\\ blue rectangle};
\node[inner sep=0pt] (part1) at (5,-2)
    {\includegraphics[width=.35\textwidth]{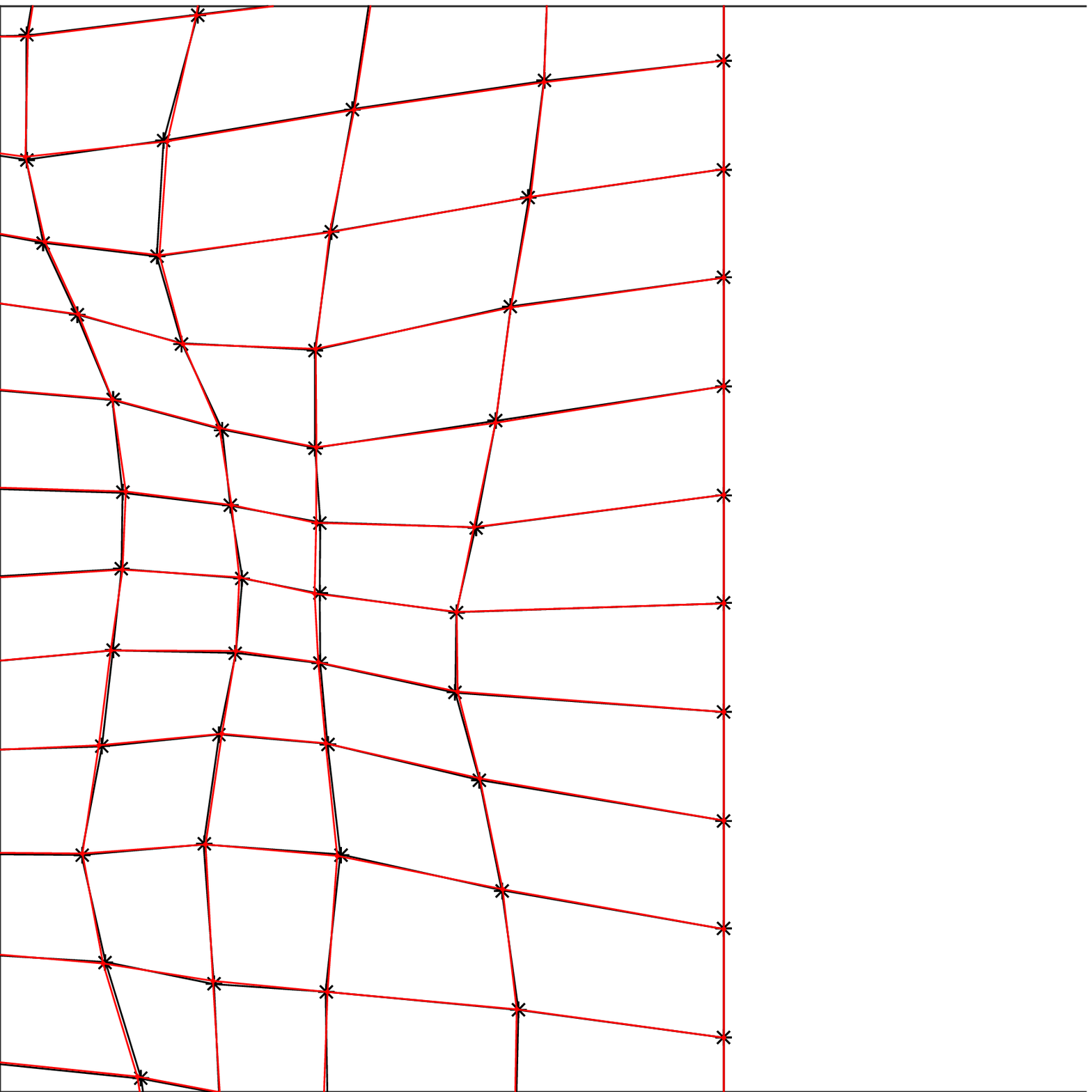}};
\node [below,align=center] at (5,-4) {(c) Enlarged view of the right\\ blue rectangle};
\draw [->,blue] (-1,1.8) to [out=90,in=180] (2.9,3);
\draw [->,blue] (2.1,0.5) to [out=0,in=180] (2.9,-2);
\end{tikzpicture}
\caption{Reconstruction of a transformation grid from its cell size and local rotation. \footnotesize{The black star dots $*$ represent $\bm{\Phi}_0$, and red dots $\cdot$ represent constructed $\bm{\Phi}_0^{(2)}$}, $65\times65$ grid size, 2.71s by a Matlab program on a general laptop.}
\end{center}
\end{figure}
\section{Averaging Two Diffeomorphisms}
We now give a numerical example about averaging two diffeomorphisms by our proposed method. We use the variational method to construct the average of diffeomorphisms $\Phi_1$ and $\Phi_2$, which are approximated by two grids, see Fig 6.(a) and(b) below. These two grids are obtained by rotating the grid $\Phi_0$ in the previous example through an angle $\theta$ in the clockwise and the counter-clockwise direction, respectively. Since the sell size does not change in rotations, and the curls of the two rotations are opposite to each other, we should expect the correct average of $\Phi_1$ and $\Phi_2$ is $\Phi_0$. We define
\[
\left\{
\begin{aligned}
f_0&=\frac 1 2(J(\Phi_1)+J(\Phi_2))\\
g_0&=\frac 1 2(curl(\Phi_1)+curl(\Phi_2)).
\end{aligned}
\right.
\]
Using the above $f_0$ and $g_0$ in the variational method, we generated the average $\Phi^*$ of $\Phi_1$ and $\Phi_2$ which is almost identical to $\Phi_0$, see Fig. 6(e,f). While the Euclidean average of $\Phi_1$ and $\Phi_2$ is shown in (c,d), which is remarkably different from $\Phi_0$.
\begin{figure}[H]
\begin{center}
\subfigure[$\Phi_1$, rotate $\Phi_0$ to the left]{\includegraphics[width=0.35\textwidth]{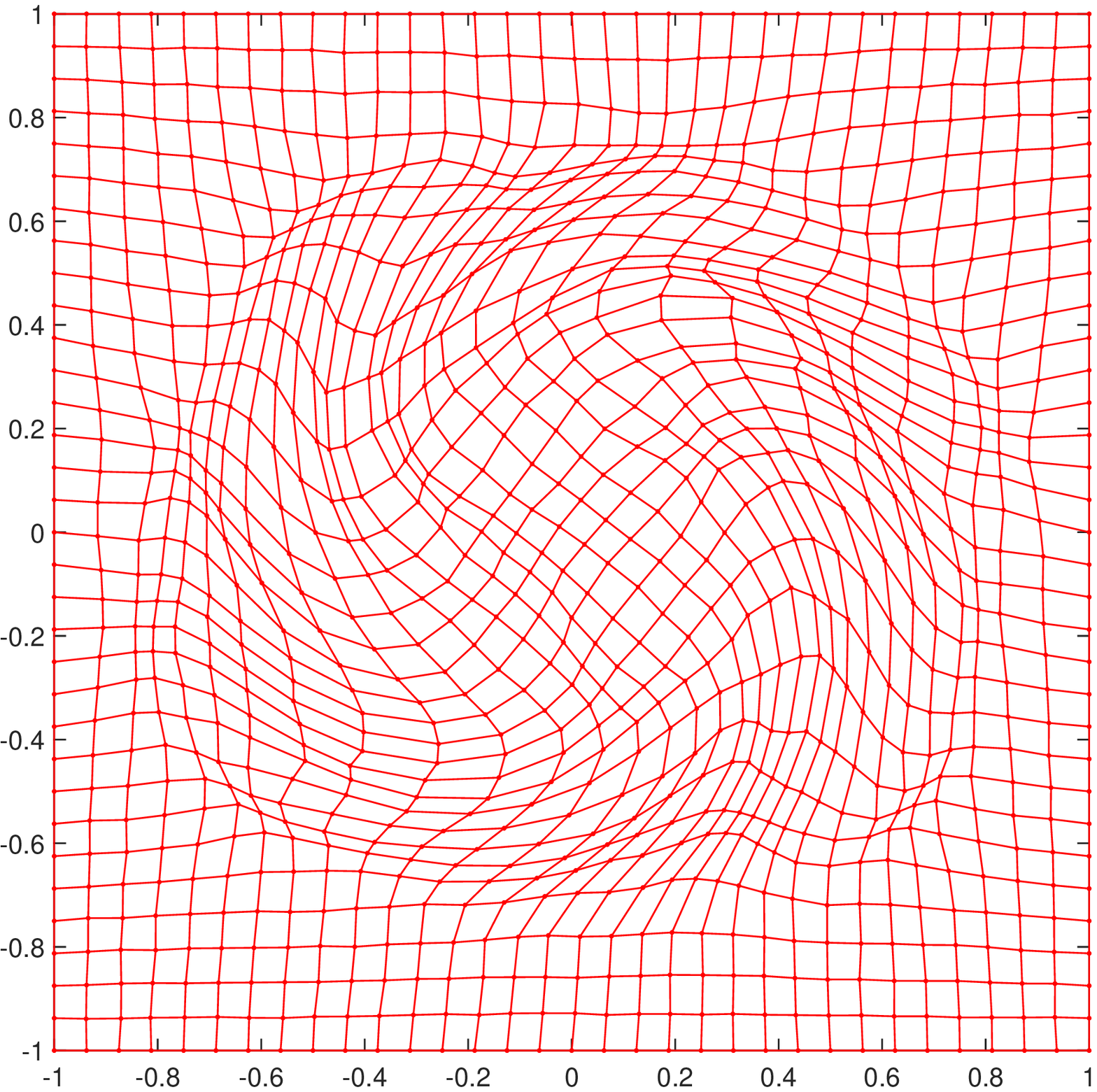}}
\subfigure[$\Phi_2$, rotate $\Phi_0$ to the right]{\includegraphics[width=0.35\textwidth]{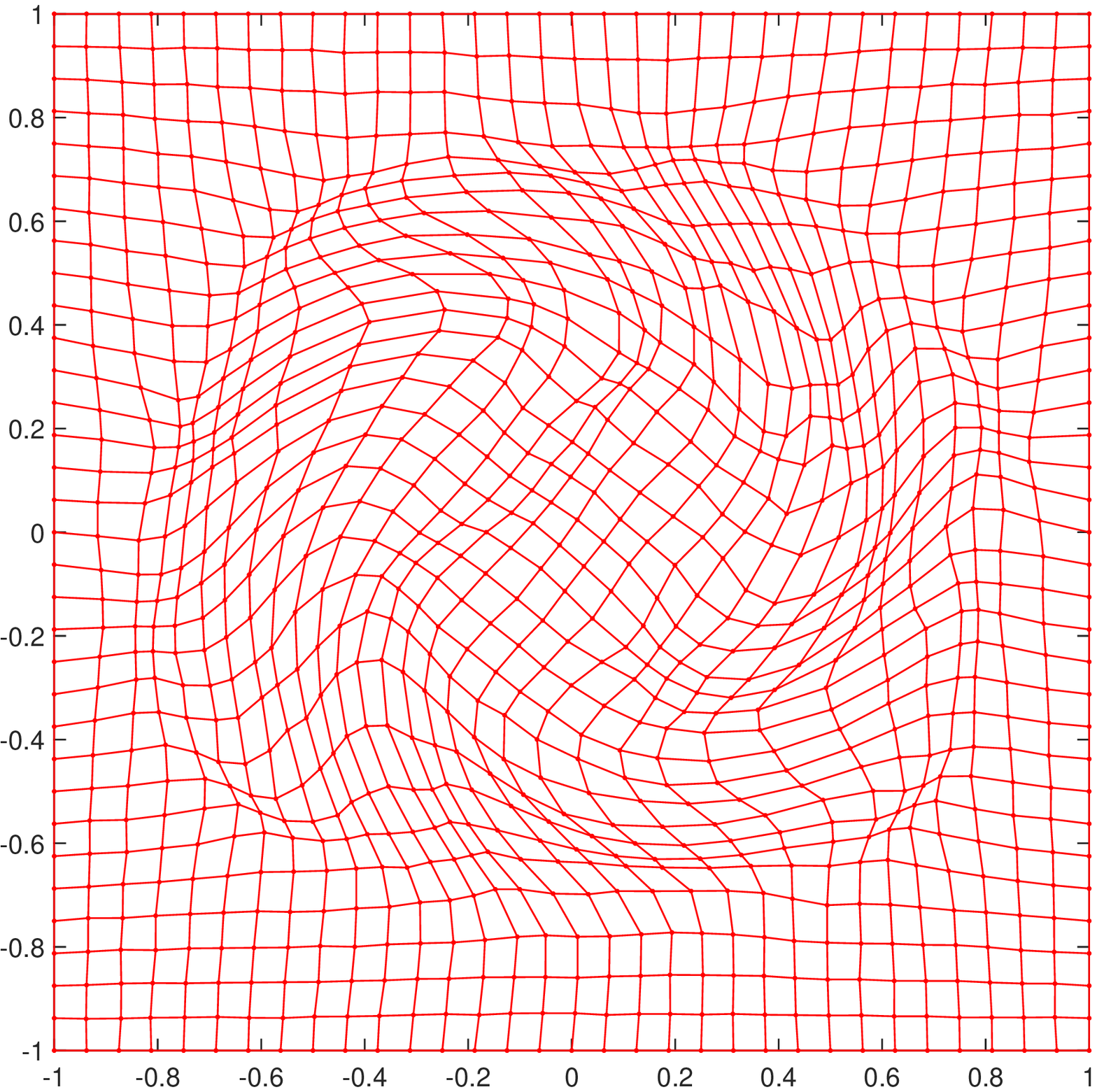}}
\subfigure[Euclidean average $\bar{\Phi}$]{\includegraphics[width=0.35\textwidth]{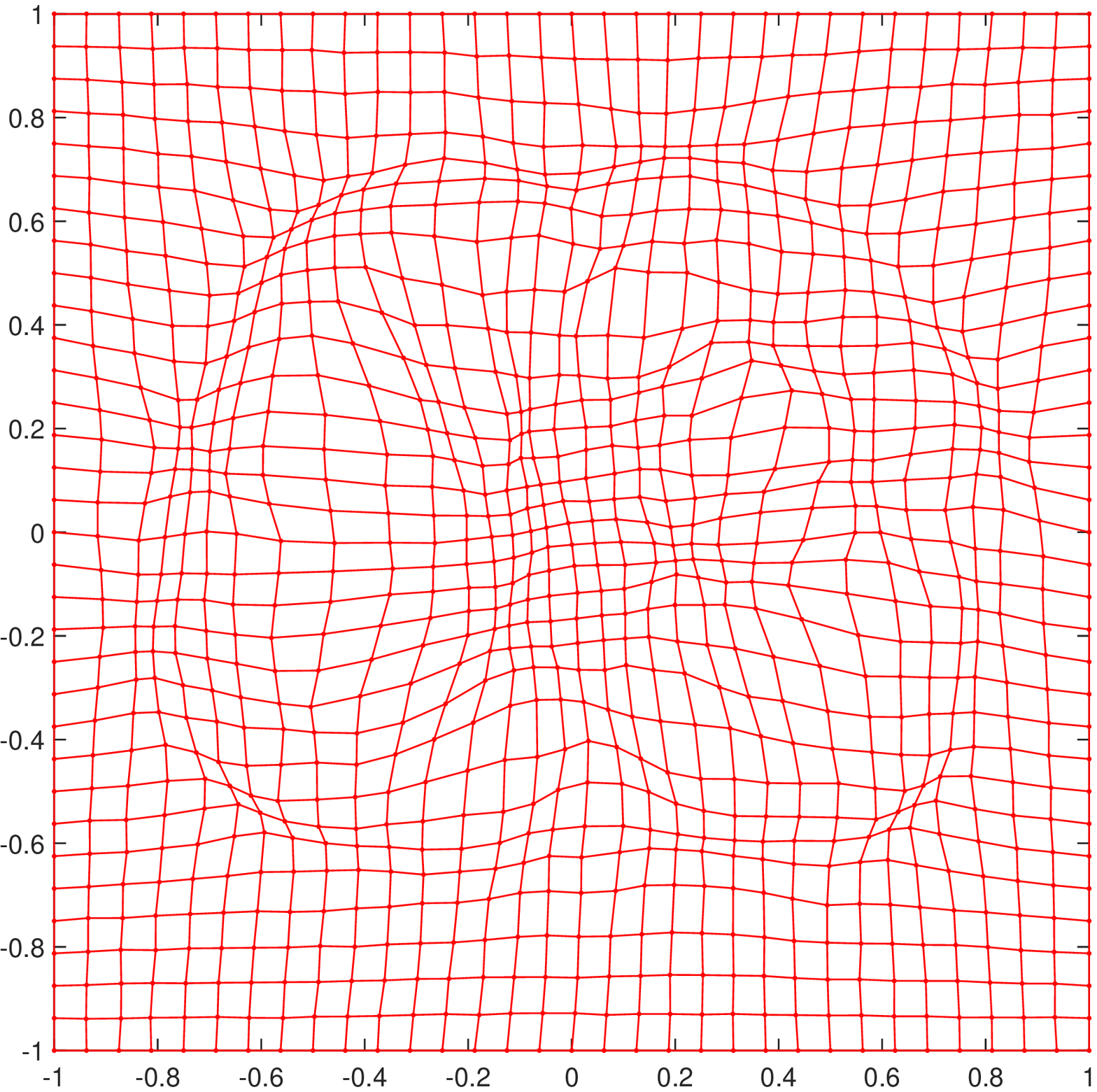}}
\subfigure[Comparison of $\bar{\Phi}$(red) and $\Phi_0$(black)]{\includegraphics[width=0.35\textwidth]{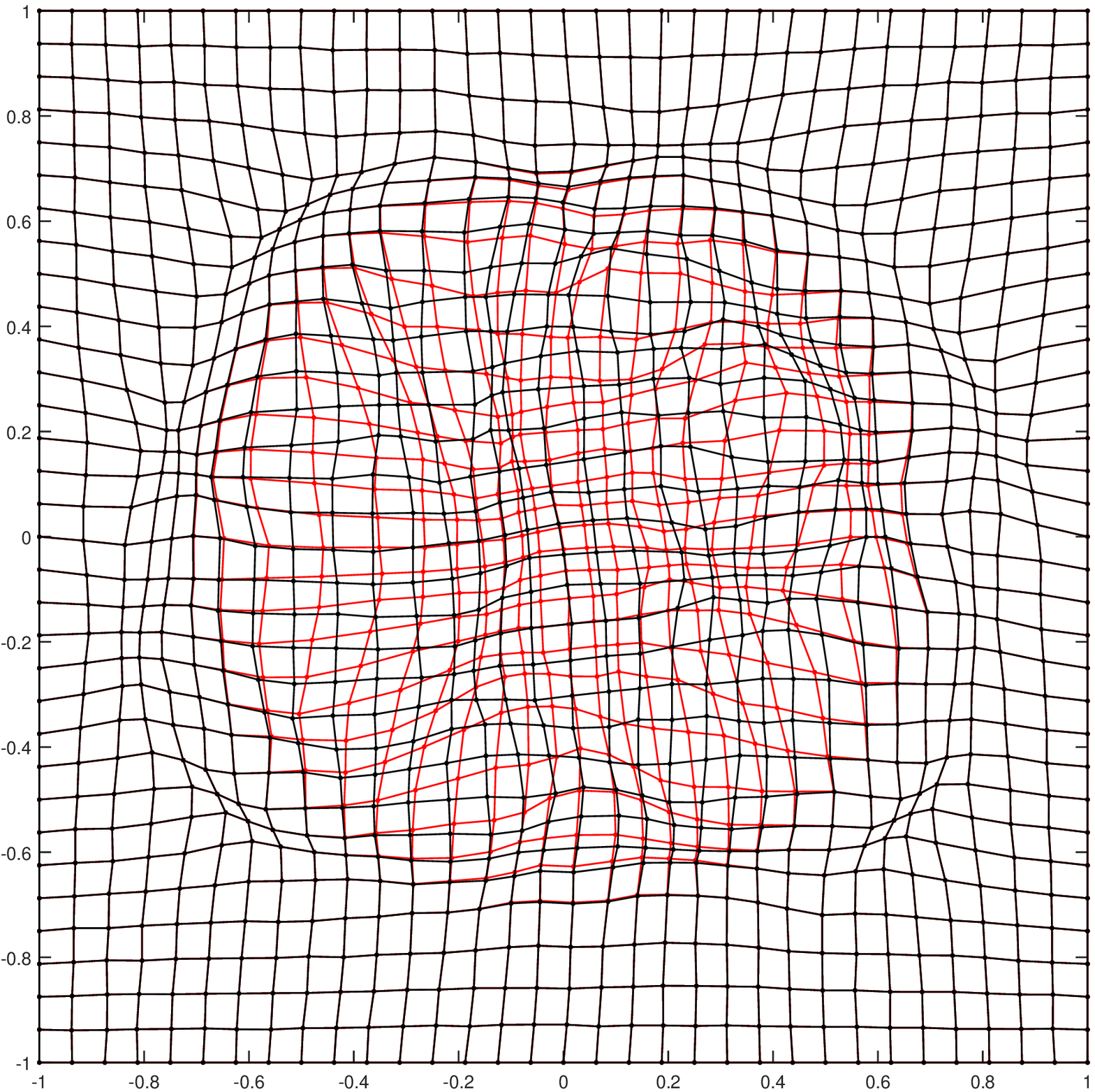}}
\subfigure[Our average $\Phi^*$]{\includegraphics[width=0.35\textwidth]{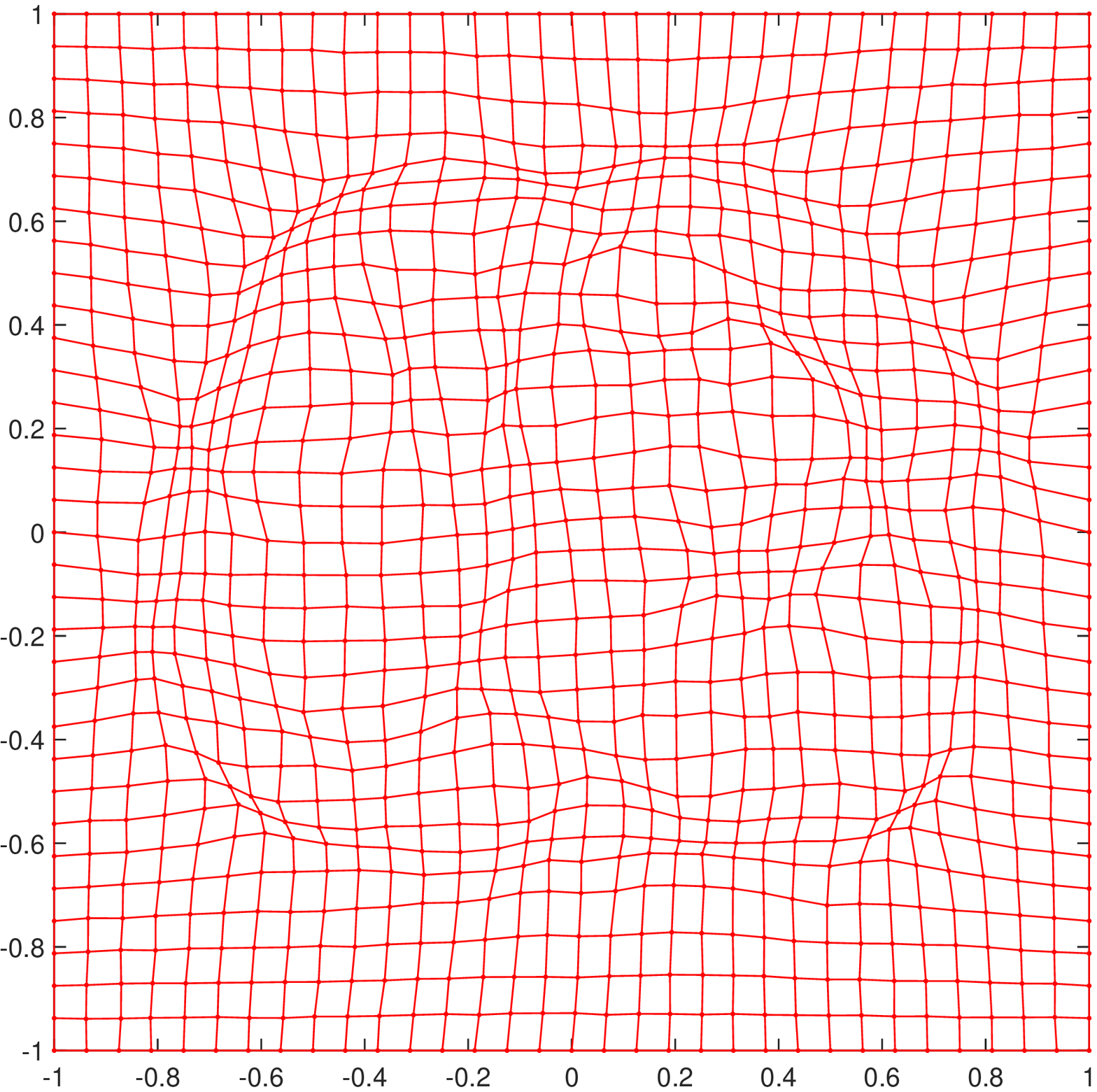}}
\subfigure[Comparison of $\Phi^*$(red) and $\Phi_0$(black)]{\includegraphics[width=0.35\textwidth]{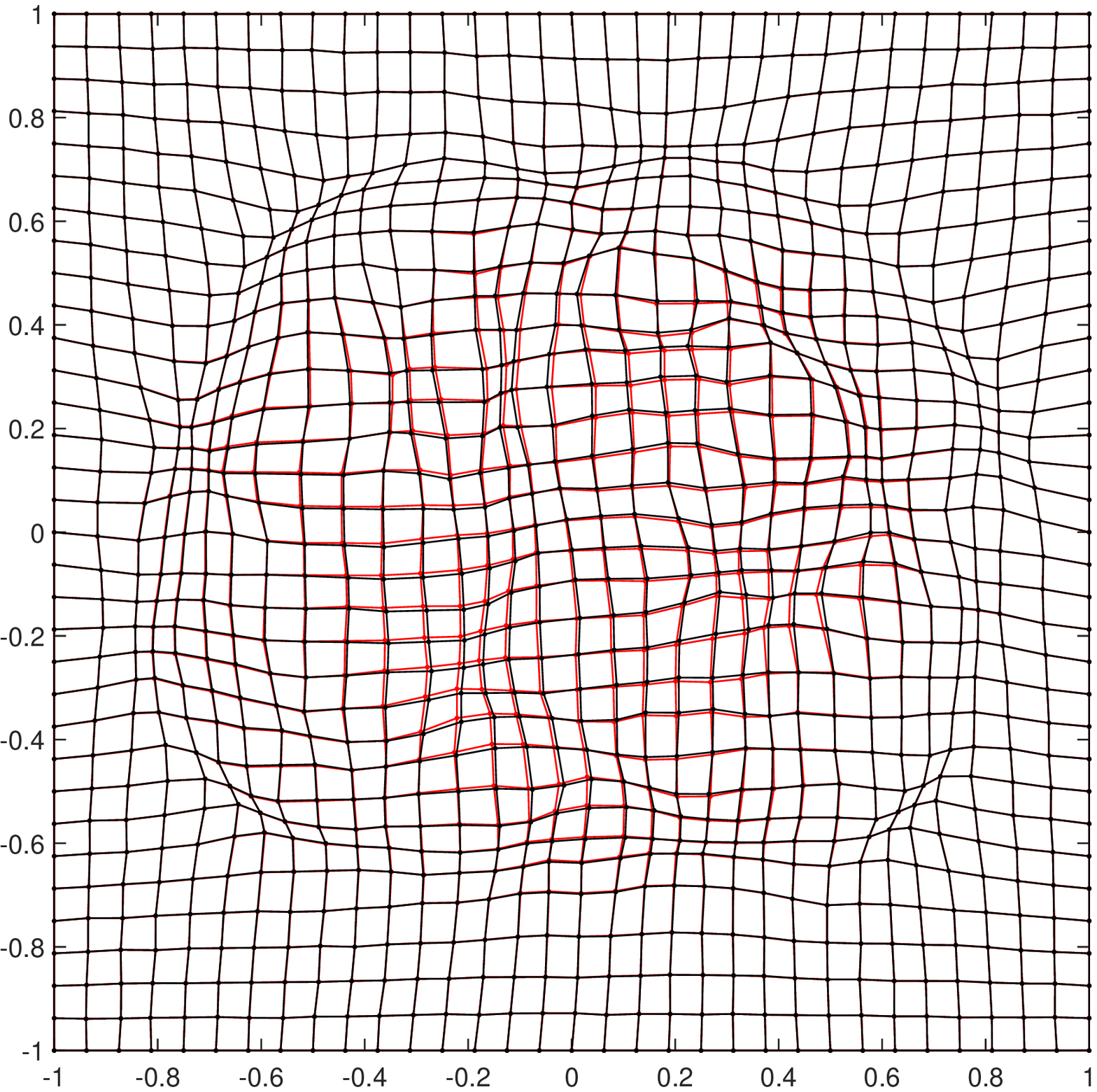}}
\caption{Averaging two diffeomorphisms by the proposed method.}
\end{center}
\end{figure}
\section{Conclusion}
In this paper, a new concept of averaging diffeomorphisms is proposed. It is based on a variational method of constructing diffeomorphisms with prescribed Jacobian determinant and curl vector. It is demonstrated by a numerical example that the method is more realistic than the Euclidean average, and the numerical algorithm is accurate and efficient. Also, the method can be extended naturally to general 3-dimensional case \cite{Chen2016}.

\end{document}